\documentclass[prd,showpacs,twocolumn,preprintnumbers,amsmath,amssymb,superscriptaddress,floatfix,nofootinbib]{revtex4-2}%

\usepackage{graphicx}
\usepackage{bm}
\usepackage{amsmath}
\usepackage{amsfonts}
\usepackage{amssymb}
\usepackage{color}
\usepackage{multirow}
\usepackage{subfigure}
\usepackage{epstopdf}
\usepackage[colorlinks, citecolor=blue,anchorcolor=red,menucolor=red, linkcolor=red,filecolor=red,runcolor=red,urlcolor=blue,frenchlinks=red]{hyperref}

\begin{document}
	\title{Hadronic decay $D^+\to\pi^+\eta\eta$ and the $a_0(980)$ and $f_0(1370)$ contributions}

	\author{Wen-Tao Lyu}
\affiliation{School of Physics, Zhengzhou University, Zhengzhou 450001, China}
	\vspace{0.5cm}
 
	\author{Le-Le Wei}\email{llwei@mails.ccnu.edu.cn}
\affiliation{Institute of Particle Physics and Key Laboratory of Quark and Lepton Physics (MOE), Central China Normal University, Wuhan, Hubei 430079, China}\vspace{0.5cm}

 	\author{De-Min Li}\email{lidm@zzu.edu.cn}
\affiliation{School of Physics, Zhengzhou University, Zhengzhou 450001, China}\vspace{0.5cm}

	\author{En Wang}\email{wangen@zzu.edu.cn}
\affiliation{School of Physics, Zhengzhou University, Zhengzhou 450001, China}\vspace{0.5cm}

\begin{abstract}	
Motivated by the resent BESIII Collaboration measurements of the process $D^+\to\pi^+\eta\eta$, we investigate this process by considering the contribution from the resonance $a_0(980)$, which is generated by $S$-wave pseudoscalar meson-pseudoscalar meson interaction, and the contribution from the intermediate scalar meson $f_0(1370)$. Our results could give a good description of the $\pi^+\eta$ and $\eta\eta$ invariant mass distributions, which implies that the contribution from the intermediate $f_0(1370)$, neglected in BESIII analysis, could play a role in this process. Furthermore, the predicted Dalitz plots are also consistent with the BESIII measurements. The more precise measurements about this process in future could shed light on the mechanism of this process and the properties of $f_0(1370)$ and $a_0(980)$.


\end{abstract}
	
	\date{\today}
	
	\maketitle
	
\section{Introduction}\label{sec1}
In the traditional quark model, hardrons are categorized into mesons which are composed of a quark-antiquark pair, and baryons which are composed of three quarks~\cite{Gell-Mann:1964ewy,Zweig:1964ruk}. Although most of the mesons could be well described within the traditional quark models, some mesons have exotic properties which are difficult to be explained. In 2003, the Belle Collaboration reported an  exotic state $X(3872)$~\cite{Belle:2003nnu}, and since that, many candidates of the exotic states were reported by experiments, and called much attentions~\cite{Chen:2022asf,Brambilla:2019esw,Meng:2022ozq,Guo:2023xyf,Liu:2024uxn,Yang:2024idy,Wang:2024jyk}. In the scalar meson sector, there are also many candidates for exotic states, for instance, $a_0(980)$, $f_0(980)$, $f_0(500)$, $f_0(1370)$, and $f_0(1710)$. As we know, the identification of the light scalar mesons is very difficult, resulting from their large decay widths, and there are many different interpretations for the structures of the scalar mesons, such as traditional $q\bar{q}$ states, multiquark states, hadronic molecules, glueballs, or the mixing of different components, etc~\cite{Close:2002zu,Amsler:2004ps,Bugg:2004xu,Klempt:2007cp,Pelaez:2015qba,Nieves:1998hp,Janssen:1994wn,Wolkanowski:2015lsa,Jaffe:1976ig,Jaffe:1999ze,Jaffe:2007id}.

Several years ago, the BESIII Collaboration has observed the process $D^+\to \pi^+\eta\eta$ using  an integrated luminosity of 2.93~fb$^{-1}$ collected
at the center-of-mass energy of 3.773~GeV and reported the branching fraction $\mathcal{B}=(2.96\pm0.24\pm0.10)\times 10^{-3}$~\cite{BESIII:2019xhl}. Later, the authors of Ref.~\cite{Ikeno:2021kzf} have studied the process $D^+\to \pi^+\eta\eta$ by considering the $a_0(980)$ dynamically generated from the pseudoscalar meson-pseudoscalar meson interaction, and predicted the signal of $a_0(980)$ in the $\pi^+\eta$ invariant mass distribution.
Recently, the BESIII Collaboration has updated the measurements of the process $D^+\to\pi^+\eta\eta$ using 20.3~fb$^{-1}$ of $e^+e^-$ collision data at the center-of-mass energy 3.773~GeV, and found a clear cusp structure in the $\pi^+\eta$ invariant mass distribution, which should be associated with the scalar $a_0(980)$~\cite{BESIII:2025yag}, as predicted in the theoretical work~\cite{Ikeno:2021kzf}. The BESIII amplitude analysis of this process suggests that the $a_0(980)$ line-shape cannot be well described by a three-channel coupled Flatt{\' e} formula for the $a_0(980)$, and could arise from the triangle loop rescattering of $D^+\to \bar{K}^*_0(1430)K^+\to a_0(980)^+\eta$ and $D^+ \to K^*_0(1430)^+ \bar{K}^0 \to a_0(980)^+ \eta$ with a significance of $5.8\sigma$~\cite{BESIII:2025yag}. However, with Eq.~(18) of Ref.~\cite{Bayar:2016ftu}, one can see that the triangle singularity of  $D^+\to \bar{K}^*_0(1430)K^+\to a_0(980)^+\eta$ appears at $M_{\pi^+\eta}=1073$~MeV, about $80$~MeV away from the nominal $a_0(980)$ mass, which implies that the triangle diagram proposed by BESIII~\cite{BESIII:2025yag} is far away from developing a triangle singularity. Indeed, the $a_0(980)$ could be dynamically generated from the $S$-wave interaction of the coupled channels $K\bar{K}$ and $\pi\eta$ within the chiral unitary approach~\cite{Oller:1997ti,Nieves:1998hp}, which has been widely used in many theoretical studies~\cite{Wang:2020pem,Xie:2014tma,Oset:2016lyh,Duan:2020vye,Zhu:2022guw,Feng:2020jvp,Wang:2022nac,Wang:2021naf,Liang:2016hmr,Lin:2021isc,Li:2025gvo,Duan:2024czu,Lyu:2024qgc,Zhang:2024myn,Liang:2019jtr}. Therefore, we need to further investigate the mechanism of the process $D^+\to\pi^+\eta\eta$ by considering the $a_0(980)$, dynamically generated from the $S$-wave pseudoscalar meson-pseudoscalar meson interaction.

In this work, we will investigate the process $D^+\to\pi^+\eta\eta$ by considering the contribution from the resonance $a_0(980)$ which is generated by $S$-wave pseudoscalar meson-pseudoscalar meson interaction. In addition, we will take into account the contribution of the light scalar $f_0(1370)$. Then, we will calculate the $\pi^+\eta$ and $\eta\eta$ invariant mass distributions in the process $D^+\to\pi^+\eta\eta$, respectively, and contrast the results with experiment with or without $f_0(1370)$. The work done here should be an incentive for a more accurate experimental analysis to be performed.

This paper is organized as follows. In Sec.~\ref{sec2}, we present the theoretical formalism of the process $D^+\to\pi^+\eta\eta$. Numerical results and discussions are shown in Sec.~\ref{sec3}. Finally, we give a short summary in the last section.

\section{Formalism}\label{sec2}

In this section, we will present the theoretical formalism. We first consider the process from the perspective that the $a_0(980)$ is dynamically generated by the $S$-wave pseudoscalar meson-pseudoscalar meson in Sec.~\ref{sec2a}, and the mechanism of the intermediate state $f_0(1370)$ is given in Sec.~\ref{sec2b}. Finally, we give the formalism for the invariant mass distributions for the process $D^+\to\pi^+\eta\eta$ in Sec.~\ref{sec2c}.

\subsection{Contribution from $a_0(980)$}\label{sec2a}

\begin{figure}[htbp]
	\centering
	
	\includegraphics[scale=0.65]{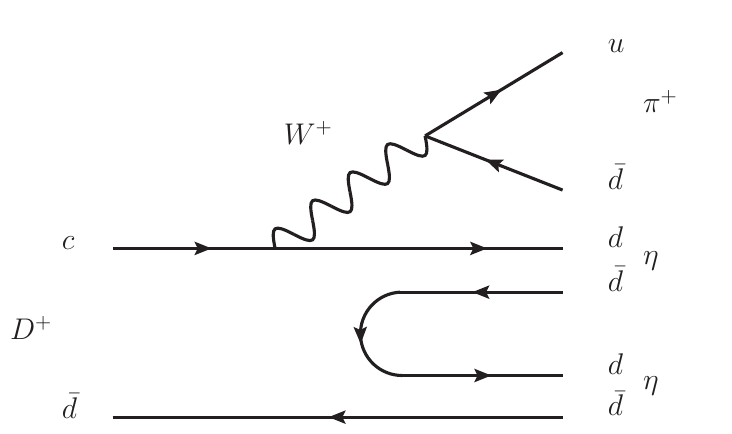}
	
	\caption{ Quark level diagram for the process $D^+\to\pi^+\eta\eta$ via the $W^+$ external emission.}\label{fig:a0-quark-1}
\end{figure}

\begin{figure}[htbp]
	\centering
	
	\includegraphics[scale=0.65]{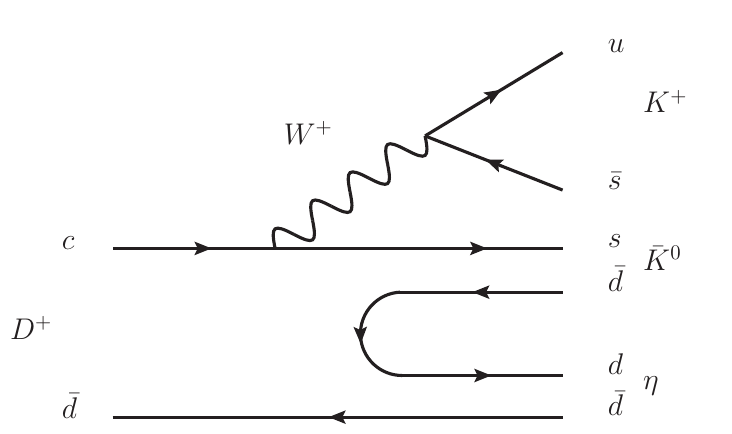}
	
	\caption{Quark level diagram for the process $D^+\to K^+\bar{K}^0\eta$ via the $W^+$ external emission.}\label{fig:a0-quark-2}
\end{figure}

Taking into account that $a_0(980)$ could be generated via $\pi^+\eta$ and $K^+\bar{K}^0$ final state interaction, as done in Refs.~\cite{Lin:2021isc,Li:2025gvo,Duan:2024czu}, we first show the quark level diagram for the $D^+\to\pi^+\eta\eta$ process via the $W^+$ external emission in Fig.~\ref{fig:a0-quark-1}. The $c$ quark from the initial $D^+$ meson weakly decays into a $W^+$ boson and a $d$ quark, then the $W^+$ boson decays into $u\bar{d}$ quark pair. The $u\bar{d}$ quark pair from the $W^+$ boson will hadronize into $\pi^+$, while the $d$ quark and the $\bar{d}$ quark of the initial $D^+$ meson, together with the antiquark-quark pair $\bar{d}d$, which is created from the vacuum with the quantum numbers $J^{PC}=0^{++}$, will hadronize into $\eta$ and $\eta$ mesons, as follow,
\begin{align}
	D^+&= c\bar{d} \nonumber\\
	&\Rightarrow W^{+}d\bar{d} \nonumber\\
	&\Rightarrow u\bar{d}d\left(\bar{d}d\right)\bar{d} \nonumber\\
	&\Rightarrow \pi^+d\left(\bar{d}d\right)\bar{d} \nonumber\\
	&\Rightarrow \frac{1}{3}\pi^{+}\eta\eta,
\end{align}
where we use the meson flavor wave functions $D^+=c\bar{d}$, $\pi^+=u\bar{d}$, and $\eta=\frac{1}{\sqrt{3}}\left(\bar{u}u+\bar{d}d-\bar{s}s\right)$ as Ref.~\cite{Close:1979bt,Miyahara:2016yyh}. 

Similar, one has another quark level diagram for the $D^+\to K^+\bar{K}^0\eta$ process via the $W^+$ external emission in Fig.~\ref{fig:a0-quark-2}. The $c$ quark from the initial $D^+$ meson weakly decays into a $W^+$ boson and an $s$ quark, then the $W^+$ boson decays into $u\bar{s}$ quark pair. The $u\bar{s}$ quark pair from the $W^+$ boson will hadronize into $K^+$, while the $s$ quark and the $\bar{d}$ quark of the initial $D^+$ meson, together with the antiquark-quark pair $\bar{d}d$ created from the vacuum, will hadronize into $\bar{K}^0$ and $\eta$ mesons, as follow,
\begin{align}
	D^+&= c\bar{d} \nonumber\\
	&\Rightarrow W^{+}s\bar{d} \nonumber\\
	&\Rightarrow u\bar{s}s\left(\bar{d}d\right)\bar{d} \nonumber\\
	&\Rightarrow K^+s\left(\bar{d}d\right)\bar{d} \nonumber\\
	&\Rightarrow \frac{1}{\sqrt{3}}K^{+}\bar{K}^0\eta,
\end{align}
where we also use the meson flavor wave functions $D^+=c\bar{d}$, $K^+=u\bar{s}$, $\bar{K}^0=s\bar{d}$, and $\eta=\frac{1}{\sqrt{3}}\left(u\bar{u}+d\bar{d}-s\bar{s}\right)$ as Ref.~\cite{Close:1979bt,Miyahara:2016yyh}.

Next, we could write all final meson-meson components after the hadronization,
\begin{align}\label{eq:lambda-dianhe}
	D^{+} \Rightarrow \frac{1}{\sqrt{3}}\eta\left( \frac{1}{\sqrt{3}}\pi^{+}\eta+K^+\bar{K}^0\right).
\end{align}

\begin{figure}
	\subfigure[]{
		\includegraphics[scale=0.65]{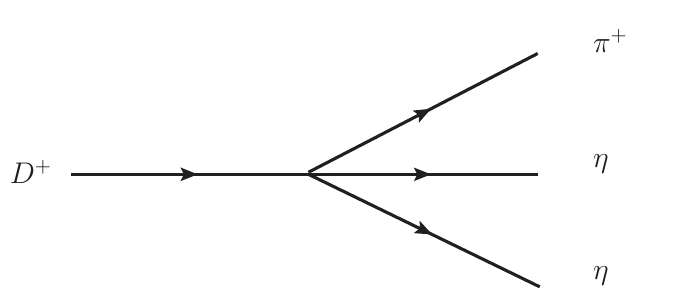}
	}
	\subfigure[]{
		\includegraphics[scale=0.65]{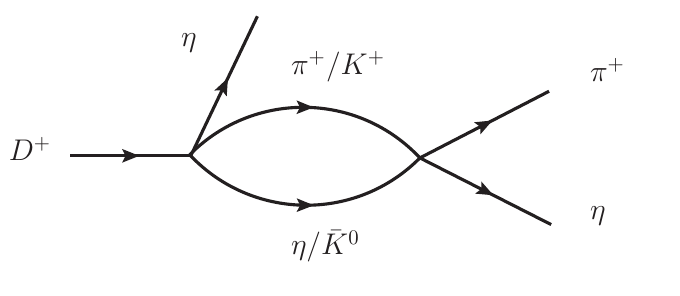}
	}
	\caption{Mechanisms of the process $D^+\to\pi^+\eta\eta$. (a) tree level, (b) the $S$-wave pseudoscalar meson-pseudoscalar meson interaction.}\label{Fig:a0-hardron}
\end{figure}

Then, the decay $D^+\to \pi^+\eta\eta$ could happen through the tree level and $S$-wave pseudoscalar meson-pseudoscalar meson interaction, which will dynamically generate the $a_0(980)$ state, as show in Figs.~\ref{Fig:a0-hardron}(a) and \ref{Fig:a0-hardron}(b), respectively. Since there are two $\eta$ mesons in the final states, one can write the decay amplitudes as follows,
\begin{eqnarray}
	\mathcal{T}^{\text{Tree}}&=&V_p h_{\pi^+\eta}, \\
	\mathcal{T}^{a_0}&=&V_p \left[ \sum_ih_iG_i(\pi^+\eta_1)t_{i\to\pi^+\eta}(\pi^+\eta_1) \right.\nonumber\\
	&& \left. +\sum_ih_iG_i(\pi^+\eta_2)t_{i\to\pi^+\eta}(\pi^+\eta_2)\right], \label{Eq:t-a0}
\end{eqnarray} 
where  $V_p$ is a global normalization constant\footnote{Here the parameter $V_{p}$ contains all dynamical factors of the weak production vertex of Figs.~\ref{Fig:a0-hardron}(a) and \ref{Fig:a0-hardron}(b). Since in this work we mainly focus on the final state interactions of this process, we assume $V_{p}$ to be constant, as done in Refs.~\cite{Zhang:2022xpf,Zhu:2022wzk,Peng:2024ive}.}, and $i=1,2$ correspond to the $\pi^+\eta$ and $K^+\bar{K}^0$ channels, respectively. The coefficients $h_i$ could be obtained from Eq.~(\ref{eq:lambda-dianhe}), 
\begin{equation}
	h_{\pi^+\eta} =\frac{1}{3}, \quad h_{K^+\bar{K}^0}=\sqrt{\frac{1}{3}}.
\end{equation}

The $G_i$ in Eq.~(\ref{Eq:t-a0}) is the loop function of the meson-meson system~\cite{Duan:2024czu},
\begin{equation}\label{loop function}
	G_i=i\int\frac{{d}^4q}{(2\pi)^4}\frac{1}{(P-q)^2-m_{1i}^2+i\epsilon}\frac{1}{q^2-m_{2i}^2+i\epsilon},
\end{equation}
where $m_{1i}$ and $m_{2i}$ are the masses of two mesons in $i$-th coupled channel, respectively, $P$ is the four-momentum of the meson-meson system, and $q$ is the four-momentum of meson in the center-of-mass frame. For the meson-meson loop function, we take the cut-off method~\cite{Guo:2005wp}, and the loop function could be written as,  
\begin{equation}
	\begin{aligned}
		G(s)=&\frac{1}{16\pi^{2}s}\left\{\sigma\left(\arctan\frac{s+\Delta}{\sigma\lambda_{1}}+\arctan\frac{s-\Delta}{\sigma\lambda_{2}}\right)\right. \\
		&\left.-\left[\left(s+\Delta\right)\ln\frac{q_{\rm max}\left(1+\lambda_{1}\right)}{m_{1}}\right.\right. \\
		&\left.\left.+\left(s-\Delta\right)\ln\frac{q_{\rm max}\left(1+\lambda_{2}\right)}{m_{2}}\right]\right\},
	\end{aligned}
\end{equation}
where
\begin{equation}
	\sigma=\left[-\left(s-(m_{1}+m_{2})^{2}\right)\left(s-(m_{1}-m_{2})^{2}\right)\right]^{1/2},
\end{equation}
\begin{equation}
	\Delta=m_{1}^{2}-m_{2}^{2},
\end{equation}
\begin{equation}
	\lambda_1=\sqrt{1+\frac{m_{1}^{2}}{q_{\max}^{2}}},~~~~\lambda_2=\sqrt{1+\frac{m_{2}^{2}}{q_{\max}^{2}}},
\end{equation}
with $s=P^2=M^2_{\pi^+\eta}$.
In order to dynamically generate $a_0(980)$ state, we adopt the cut-off parameter $q_\mathrm{max}=600$~MeV as used in Refs.~\cite{Lin:2021isc,Li:2025gvo,Duan:2024czu}.

The $t_{i\to\pi^+\eta}$ in Eq.~(\ref{Eq:t-a0}) are the transition amplitudes of the coupled channels, and $t_{i\to\pi^+\eta}$ could be obtained through the Bethe-Salpeter equation,
\begin{equation}\label{BS}
	T=[1-VG]^{-1}V.
\end{equation}
In our calculation, we take into account two coupled channels $\pi^+\eta$ and $K^+\bar{K}^0$. The transition potential $V_{ij}$ is obtained from Refs.~\cite{Lin:2021isc,Duan:2024czu},
\begin{align}
	&V_{\pi^+\eta\to\pi^+\eta}=-\frac{2m_{\pi}^2}{3f^2},\nonumber\\
	&V_{K^{+}\bar{K}^{0}\to K^{+}\bar{K}^{0}}=-\frac{s}{4f^{2}}, \nonumber\\
	&V_{K^{+}\bar{K}^{0}\to\pi^{+}\eta}=-\frac{3s-2m_K^2-m_\eta^2}{3\sqrt{3}f^2},
\end{align}
where the decay constant $f=f_{\pi}=93$~MeV. $m_\pi$  and $m_K$ represent the averaged masses of the pion and kaon.

\subsection{Contribution from $f_0(1370)$}\label{sec2b}

\begin{figure}[htbp]
	\centering
	
	\includegraphics[scale=0.65]{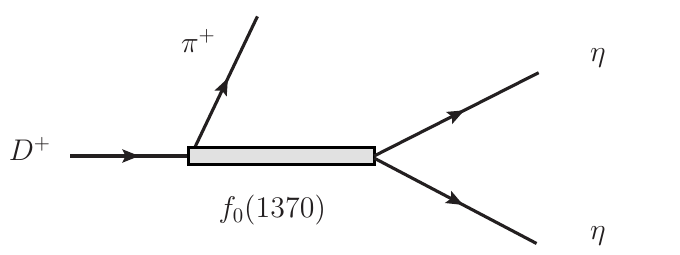}	
	\caption{Mechanism of the intermediate $f_0(1370)$.}\label{fig:f0-hardon}
\end{figure}

According to BESIII measurements of the process $D^+\to \pi^+\eta\eta$, on can find that the $\eta\eta$ invariant mass distribution is significantly different with the phase space distribution in the region $1300\sim 1500$~MeV~\cite{BESIII:2025yag}. Considering that the scalar meson $f_0(1370)$ could decay into the $\eta\eta$~\cite{ParticleDataGroup:2024cfk},  there could be the contribution from the $f_0(1370)$ state. Therefore, we can also consider the contribution from $f_0(1370)$, and the mechanism of the process $D^+\to\pi^+f_0(1370)\to\pi^+\eta\eta$ is depicted in Fig.~\ref{fig:f0-hardon}. We can write the amplitude for the intermediate $f_0(1370)$ state,
\begin{equation}
	\mathcal{T}^{f_0(1370)}=V_p\dfrac{\alpha M_{f_0(1370)}\Gamma_{f_0(1370)}}{M_{\eta\eta}^2-M_{f_0(1370)}^2+iM_{f_0(1370)}\Gamma_{f_0(1370)}},
\end{equation}
where the $\alpha$ is the relative strength of the contribution form the intermediate $f_0(1370)$ with respect to the one of $a_0(980)$. Here the $M_{f_0(1370)}$ and $\Gamma_{f_0(1370)}$ are the mass and width of the $f_0(1370)$. Since the $f_0(1370)$ is a broad state, and its mass and width have not yet been establish~\cite{ParticleDataGroup:2024cfk}, there are also many studies about the nature of the $f_0(1370)$~\cite{Bugg:2007ja,Ochs:2013gi,Molina:2008jw,Geng:2008gx}.


\subsection{Invariant Mass Distribution}\label{sec2c}

We can write the total amplitude of $D^+\to\pi^+\eta\eta$ reaction as,
\begin{equation}\label{Eq:t-total}
	\mathcal{T}^{\text{Total}}=\mathcal{T}^{\text{Tree}}+\mathcal{T}^{a_0}+\mathcal{T}^{f_0(1370)}e^{i\phi},
\end{equation}
where the phase angle $\phi$ is free parameter. Applying the standard formula of the three-body decay width from RPP~\cite{ParticleDataGroup:2024cfk}, we have
\begin{equation}
	\frac{d^2\Gamma}{d M_{\pi^+\eta_1} d M_{\eta_1\eta_2}}=\frac{1}{(2\pi)^3}\dfrac{M_{\pi^+\eta_1}M_{\eta_1\eta_2}}{8M_{D^+}^3}|\mathcal{T^{\rm Total}}|^2.
\end{equation}
Given a specific value of the invariant mass $M_{12}$, the corresponding range for the invariant mass $M_{23}$ is determined according to the RPP~\cite{ParticleDataGroup:2024cfk},
\begin{align}
	&\left(m_{23}^2\right)_{\min}=\left(E_2^*+E_3^*\right)^2-\left(\sqrt{E_2^{* 2}-m_2^2}+\sqrt{E_3^{* 2}-m_3^2}\right)^2, \nonumber\\
	&\left(m_{23}^2\right)_{\max}=\left(E_2^*+E_3^*\right)^2-\left(\sqrt{E_2^{* 2}-m_2^2}-\sqrt{E_3^{* 2}-m_3^2}\right)^2, \label{eq:limit}
\end{align}
where $E_2^*$ and $E_3^*$ are the energies of particles 2 and 3 in the $M_{12}$ rest frame, respectively,
\begin{align}
	E_{2}^{*}&=\frac{M_{12}^{2}-m_{1}^{2}+m_{2}^{2}}{2M_{12}}, \nonumber\\
	E_{3}^{*}&=\frac{M_{D^{+}}^{2}-M_{12}^{2}-m_{3}^{2}}{2M_{12}}.
\end{align}
where $m_1$, $m_2$, and $m_3$ denote the masses of particles 1, 2, and 3,  respectively. The masses and widths of the particles are taken from the RPP~\cite{ParticleDataGroup:2024cfk}. Permutation of the indices allows us to evaluate all three mass distributions, using $M_{12}$, $M_{23}$ as independent variables, and the property $M_{12}^2+M_{13}^2+M_{23}^2=M_{D^+}^2+m_{\pi^+}^2+m_\eta^2+M_\eta^2$ to get $M_{13}$ from them.

It is notable that there are two $\eta$ mesons in this process so that there have $\pi^+\eta_1$ and $\pi^+\eta_2$ invariant mass distributions. In order to better compare with the experimental data, we take,
\begin{equation}
	\frac{d\Gamma}{dM_{\pi^+\eta}}=\frac{d\Gamma}{dM_{\pi^+\eta_1}}+\frac{d\Gamma}{dM_{\pi^+\eta_2}}.
\end{equation}

\section{Results and Discussions}\label{sec3}

\begin{table*}[htpb]
	\begin{center}	
		\caption{Fitted parameters of this work. The mass of $f_0(1370)$ is in units of MeV.}
		\begin{tabular}{cccccc}
			\hline\hline
		   Parameters \qquad\quad& $V_P$\qquad\quad& $\alpha $\qquad\quad&$M_{f_0(1370)}$\qquad\quad&$\phi$\qquad\quad&$\chi^2/d.o.f.$\\
			\hline
		  without $f_0(1370)$ \qquad\quad&  $1026.9\pm8.5$\qquad\quad& /\qquad\quad& /\qquad\quad& /\qquad\quad& $1.10$  \\
            with $f_0(1370)$ \qquad\quad& $1024.2\pm29.7$ \qquad\quad& $0.07\pm0.01$\qquad\quad& $1331.5\pm95.1$\qquad\quad& $(1.22\pm0.26)\pi$\qquad\quad& $0.96$  \\
            
			\hline\hline
		\end{tabular}\label{tab:fitpar}
	\end{center}
\end{table*}

\begin{figure}
	\subfigure[]{
		\includegraphics[scale=0.6]{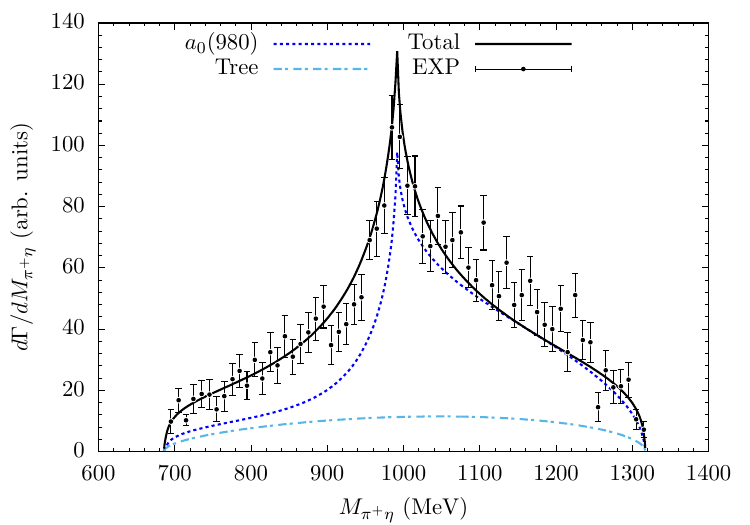}
	}
	\subfigure[]{
		\includegraphics[scale=0.6]{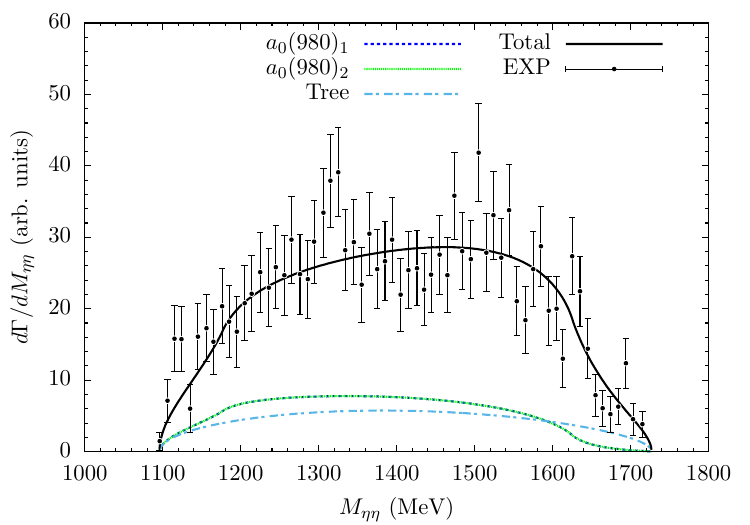}
	}
	\caption{$\pi^+\eta$~(a) and $\eta\eta$~(b) invariant mass distributions of the $D^+\to\pi^+\eta\eta$ reaction without the contribution of $f_0(1370)$.} \label{Fig:only-a0}
\end{figure}

\begin{figure}
	\subfigure[]{
		\includegraphics[scale=0.6]{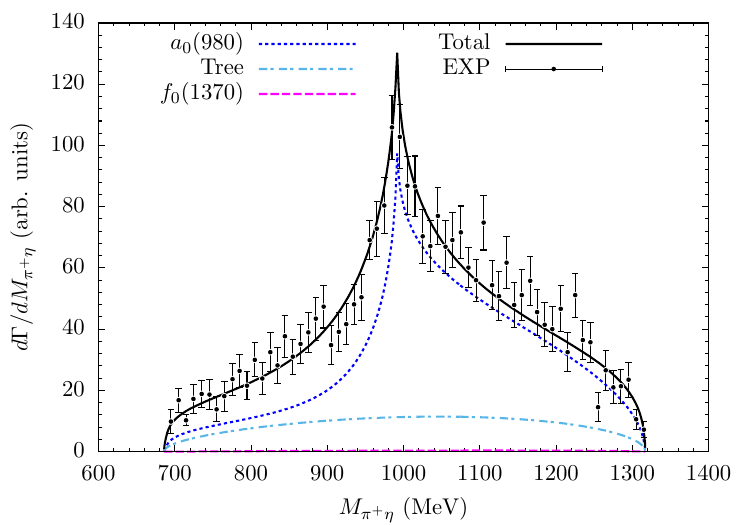}
	}
	\subfigure[]{
		\includegraphics[scale=0.6]{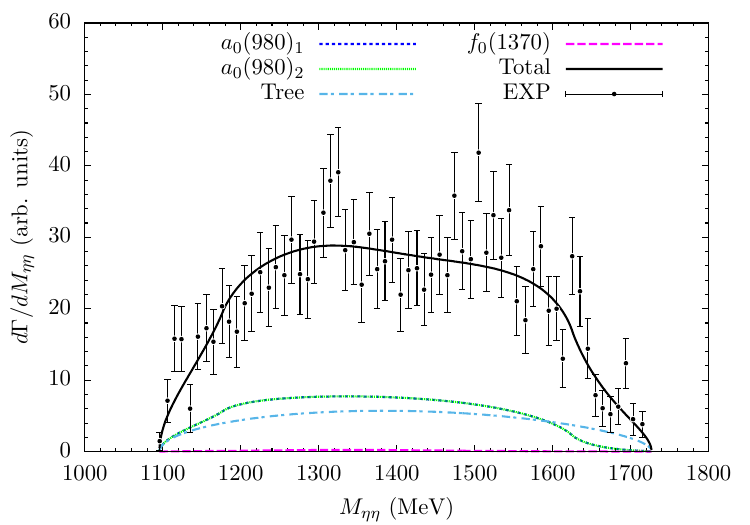}
	}
	\caption{$\pi^+\eta$~(a) and $\eta\eta$~(b) invariant mass distributions of the $D^+\to\pi^+\eta\eta$ reaction with the contribution of $f_0(1370)$.} \label{Fig:a0+f0}
\end{figure}

\begin{figure}[htbp]
	\centering
	
	\includegraphics[scale=0.80]{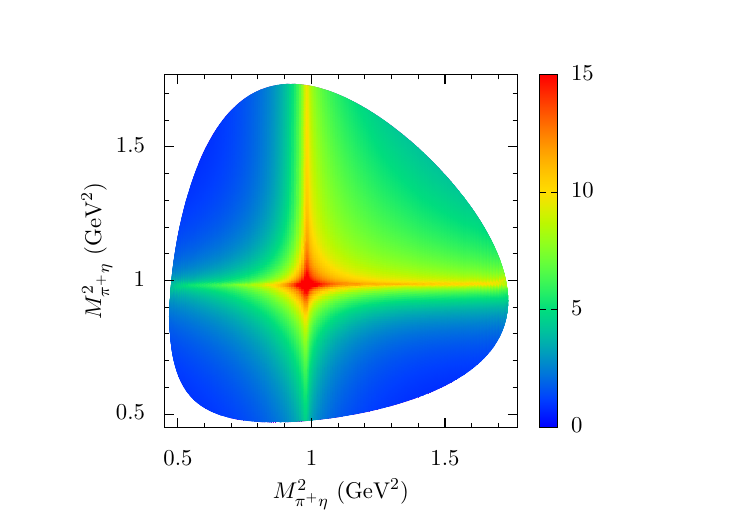}
	
	\caption{Dalitz plot of ``$M^2_{\pi^+\eta}$" vs. ``$M^2_{\pi^+\eta}$" for the double differential decay width $d^2\Gamma/dM_{\pi^+\eta}dM_{\pi^+\eta}$ of the process $D^+\to\pi^+\eta\eta$.}\label{fig:Dalitz1}
\end{figure}

\begin{figure}[htbp]
	\centering
	
	\includegraphics[scale=0.80]{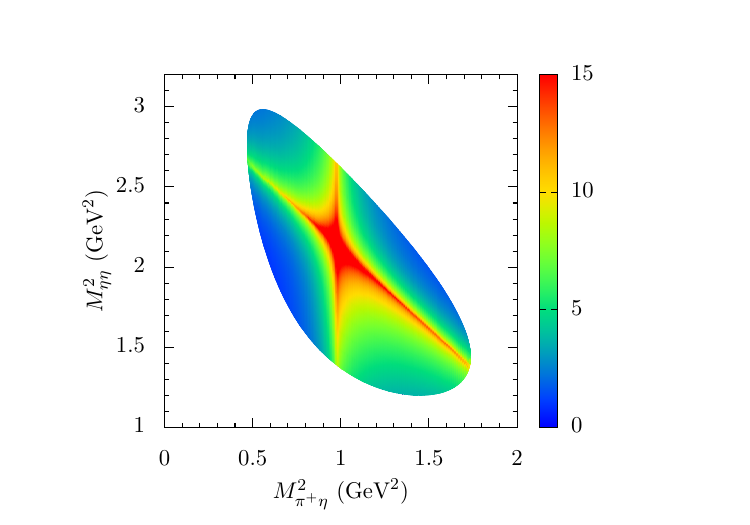}
	
	\caption{Dalitz plot of ``$M^2_{\pi^+\eta}$" vs. ``$M^2_{\eta\eta}$" for the double differential decay width $d^2\Gamma/dM_{\pi^+\eta}dM_{\eta\eta}$ of the process $D^+\to\pi^+\eta\eta$.}\label{fig:Dalitz2}
\end{figure}

As we discussed above, we can get $d\Gamma/dM_{12}$ by integrating $d^2\Gamma/(dM_{12}dM_{23})$ over $M_{23}$ with the limits of Eq.~(\ref{eq:limit}). Firstly, we have only considered the contribution from $a_0(980)$ state for comparing with the BESIII measurements~\cite{BESIII:2025yag}. It is worth mentioning that, in this case, there is only one free parameter, i.e. the global normalization constant $V_p$. The parameter of the fit is given in Table~\ref{tab:fitpar}, and we have $\chi^2/d.o.f.=137.93/(126-1)=1.10$. We show our results of the $\pi^+\eta$ invariant mass distributions in Fig.~\ref{Fig:only-a0}(a). The blue-dotted curve shows the contributions from $a_0(980)$, which includes the contributions by exchanging two $\eta$. The blue-dashed curve shows the contribution from tree level. The black-solid curve shows the total contribution. One can find a significant cusp structure around 980~MeV in the $\pi^+\eta$ invariant mass distribution, which could be associated with the $a_0(980)$ state. However, one can see that, in the 1000$\sim$1200~MeV region, our results are lower than the BESIII measurements~\cite{BESIII:2025yag}, which implies that there could be some contributions from other intermediate states. On the other hand, we show the $\eta\eta$ invariant mass distribution in Fig.~\ref{Fig:only-a0}(b). The blue-dotted curve shows the contribution from $a_0(980)_1$ ($\pi^+\eta_1$) and  the green-solid curve shows the contribution from $a_0(980)_2$ ($\pi^+\eta_2$). The blue-dashed curve shows the contribution from tree level. The black-solid curve shows the total contribution. One can also find that, in the whole energy region, our results are in fair agreement with the BESIII measurements~\cite{BESIII:2025yag}.

Furthermore, we have considered the contribution from the intermediate $f_0(1370)$ state, which can decay into $\eta\eta$ in $S$-wave. In this case, we have considered the $f_0(1370)$ state and the effect of interference between $f_0(1370)$ and other terms. Since the BESIII measurements have larger uncertainties, we fix the width of $f_0(1370)$ $\Gamma_{f_0(1370)}=350$~MeV in order to reduce the number of the free parameters. In our calculation, there are four free parameters, (1) the strength $V_p$ for the contribution from the tree diagram and the $a_0(980)$; (2) the relavtive strength $\alpha$ for the contribution from $f_0(1370)$; (3) the mass of the $f_0(1370)$ $M_{f_0(1370)}$; (4) the interference phase angle $\phi$. The parameters of the fit are given in Table~\ref{tab:fitpar}. One can find that the fitted $\chi^2/d.o.f.=117.34/(126-4)=0.96$ is smaller than the results (1.10) without the contribution from the intermediate $f_0(1370)$.

We present our results of the $\pi^+\eta$ invariant mass distributions in Fig.~\ref{Fig:a0+f0}(a). The blue-dotted curve shows the contributions from $a_0(980)$. The blue-dashed curve shows the contribution from tree level. The black-solid curve shows the total contribution. One can see that, in the 1000$\sim$1200~MeV region, the results considering the contribution from $f_0(1370)$ are in good agreement with the BESIII measurements~\cite{BESIII:2025yag}. Meanwhile, we show the $\eta\eta$ invariant mass distribution in Fig.~\ref{Fig:a0+f0}(b). The blue-dotted curve shows the contributions from $a_0(980)_1$ ($\pi^+\eta_1$). The green-solid curve shows the contributions from $a_0(980)_2$  ($\pi^+\eta_2$).  The blue-dashed curve shows the contribution from tree level. The black-solid curve shows the total contribution. One can find that, although the contribution from the $f_0(1370)$ is small, the $\eta\eta$ invariant mass distribution with the interference effect of $f_0(1370)$ is significantly different with the previous results shown in Fig.~\ref{Fig:only-a0}(b). 
Indeed, the BESIII measured $\pi^+\eta$ and $\eta\eta$ still have large fluctuation, the more precise measurments in future could be used to constrain the the contribution from the $f_0(1370)$, and determine the mass and width of $f_0(1370)$.

Finally, we show the Dalitz plots of `$M^2_{\pi^+\eta}$' vs. `$M^2_{\pi^+\eta}$' and `$M^2_{\pi^+\eta}$' vs. `$M^2_{\eta\eta}$' with the contribution from $f_0(1370)$ state in Fig.~\ref{fig:Dalitz1} and Fig.~\ref{fig:Dalitz2}, respectively. One can find the clear signal of the $a_0(980)$ around $M^2_{\pi^+\eta}=0.96~\text{GeV}^2$. 

\section{Conclusions}
Recently, the BESIII Collaboration has measured the process $D^+\to\pi^+\eta\eta$ using 20.3 fb$^{-1}$ of $e^+e^-$ collision data at the center-of-mass energy 3.773~GeV, and found a clear cusp structure in the $\pi^+\eta$ invariant mass distribution, which should be associated with the scalar $a_0(980)$~\cite{BESIII:2025yag}. The BESIII amplitude analysis of this process suggests that the $a_0(980)$ line-shape cannot be well described by the three-channel coupled Flatt{\' e} formula for the $a_0(980)$, and could arise from the triangle loop rescattering of $D^+\to \bar{K}^*_0(1430)K^+\to a_0(980)^+\eta$ with a significance of $5.8\sigma$~\cite{BESIII:2025yag}. 
However, the triangle singularity of  $D^+\to \bar{K}^*_0(1430)K^+\to a_0(980)^+\eta$ appears at $M_{\pi^+\eta}=1073$~MeV, about $80$~MeV away from the nominal $a_0(980)$ mass, which implies that the triangle diagram proposed by BESIII~\cite{BESIII:2025yag} is far away from developing a triangle singularity.

According to BESIII measurements of the process $D^+\to \pi^+\eta\eta$, on can find that the $\eta\eta$ invariant mass distribution is significantly different from the phase space distribution in the region $1300\sim 1500$~MeV~\cite{BESIII:2025yag}. Considering that the scalar meson $f_0(1370)$ could decay into $\eta\eta$~\cite{ParticleDataGroup:2024cfk},  there could be a contribution from the $f_0(1370)$ state.

Therefore, in this work, we have investigated this process by considering the contribution from the resonance $a_0(980)$, which is generated by $S$-wave pseudoscalar meson-pseudoscalar meson interaction within the chiral unitary approach, and the contribution from the intermediate scalar meson $f_0(1370)$. 

One can find the our results including the contribution from the intermediate $f_0(1370)$, neglected in BESIII analysis, give a good description of the $\pi^+\eta$ and $\eta\eta$ invariant mass distributions, which supports the molecular nature of the $a_0(980)$, and implies that the $f_0(1370)$ plays a sizeable role. Since the properties of the scalar $f_0(1370)$ have not been well established and the BESIII data have large fluctuation, the more precise measurements about this process by Belle II and the proposed STCF could shed light on the mechanism of this process and the nature of $f_0(1370)$.

\section*{Acknowledgments}
We would like to acknowledge the fruitful discussions with Profs. Eulogio Oset and Jia-Jun Wu. 
This work is supported by the National Key R\&D Program of China (No. 2024YFE0105200),  the Natural Science Foundation of Henan under Grant No. 232300421140, the National Natural Science Foundation of China under Grant No. 12475086 and No. 12192263. This work is supported by Zhengzhou University Young Student Basic Research Projects (PhD students) under Grant No. ZDBJ202522.

\end{document}